\documentclass[12pt]{iopart}

\usepackage{latexsym}
\usepackage{graphics}
\usepackage{amssymb}

\begin{document}

\title{Determination of the species generated in atmospheric-pressure laser-induced plasmas by mass spectrometry techniques}

\author{F. Valle$^1$, C. Salgado$^1$, J. I. Api\~naniz$^1$, A. V. Carpentier$^{1*}$, M. S\'anchez Albaneda$^1$, L. Roso$^1$, C. Raposo$^2$, C. Padilla$^3$, A. Peralta Conde$^1$}
\address{$^1$Centro de L\'aseres Pulsados (CLPU), Parque Cient\'{\i}fico, 37185 Villamayor, Salamanca, Spain.}
\address{$^2$Universidad de Salamanca, 37008 Salamanca, Spain.}
\address{$^3$Iberdrola Ingenier\'ia y Construcci\'on, Proyectos de I+D+i (DITE-IGEN-CIPA) 28050 Madrid, Spain}
\ead{$^*$avazquez@clpu.es}
\vspace{10pt}
\begin{indented}
\item[]December 2015
\end{indented}

\begin{abstract}
We present temporal information obtained by mass spectrometry techniques about the evolution of plasmas generated by laser filamentation in air. The experimental setup used in this work allowed us to study not only the dynamics of the filament core but also of the energy reservoir that surrounds it. Furthermore, valuable insights about the chemistry of such systems like the photofragmentation and/or formation of molecules were obtained. The interpretation of the experimental results are supported by PIC (particle in cell) simulations.
\end{abstract}

\maketitle

\section{Introduction}
During the last decades the non-stoping development of laser technology has allowed the discovery of new unexpected phenomena that have attracted a considerable attention not only for their scientific interest, but also for their possible practical applications. Nowadays lasers are the basis of tools and techniques that permit us to study nature with a level of detail not even considered a decade ago.  One paradigmatic example of new phenomena intrinsically linked with the development of the technology are laser induced plasmas like for example plasmas induced by laser filamentation. This kind of plasmas have a special interest because the parameters are fundamentally limited, e.g., density and temperature, by the characteristics of the laser propagation in the medium. Filamentation has its origin in the balance between two different effects: self focusing of the laser pulse when it propagates through a positive Kerr medium and defocusing due to the properties of the plasma created by the concentration of energy in space and time. As a consequence the propagation is "collimated" in a channel leaving a plasma in a delimited region of the medium once the radiation ceases. Filamentation in solids was first observed by Hercher in 1964 \cite{Hercher64} not so long after the first laser was successfully fired by T.H. Maiman in 1960; but it was not until the development of intense femtosecond pulsed lasers that filamentation was observed in the atmosphere \cite{Mourou95}.  Since then many different groups have devoted their efforts to study the process of filamentation  itself and the possible applications of the plasma channel induced by the laser field (see for example \cite{Mysyrowicz07,Kasparian08,Chin11,Chin10,Trapani06, Sunchu15} and references therein). If we concentrate in the field of filamentation on atmospheric Sciences some of the on-going applications are the use of laser filaments to control lighting \cite{Wolf08,Schroeder05,Wolf08_2,Miyazaki99} and rainfalls \cite{Schroeder05,Wolf10_3,Wolf10_4}, the guiding of high power microwaves using the ionisation channel created by the filament \cite{Kieffer08,Kandidov07},  or the remote sensing of pollutants in air using molecular fluorescence and LIDAR (Laser Illuminated Detection And Ranging) \cite{Chin10,Dubois09,Milonni09}.  It is important to recognise that for a successful implementation of such techniques it is required a complete understanding of the process of laser filamentation and, more importantly, of the subsequent plasma dynamics. Thus, questions like which species are formed inside the plasma channel, at what time after the formation of the filament new species are generated, or at what time after the pulse the plasma density becomes negligible and the species can move freely, need to be addressed to provide predictable and reliable models not only of the filamentation process itself but also of plasmas at atmospheric pressure conditions. So far different groups have dealt with these problems mainly using spectroscopic techniques like Raman spectroscopy (see for example \cite{Levis08,Romanov09,Chin01}) providing useful information about the process dynamics. However, these analysis rely on the correct identification of the fluorescence lines of the plasma which is usually a difficult and arduous task. 

In this work we do propose a complete different approach. Getting advantage of the high ionization yields induced by lasers, we have used a modified version of a Time of Flight spectrometer (TOF) developed in our laboratory where the sample does not need to be isolated from the circumvent medium for the ionization process. Thus, we have obtained detailed information on the different ion species generated within the plasma, in the inner core as well as the surroundings, at atmospheric pressure conditions, as well as on the relaxation mechanisms related with the plasma dynamics like for example molecular recombination. 

\section{Experimental setup}

\begin{figure}[ht!]
\resizebox{1\textwidth}{!}{\includegraphics{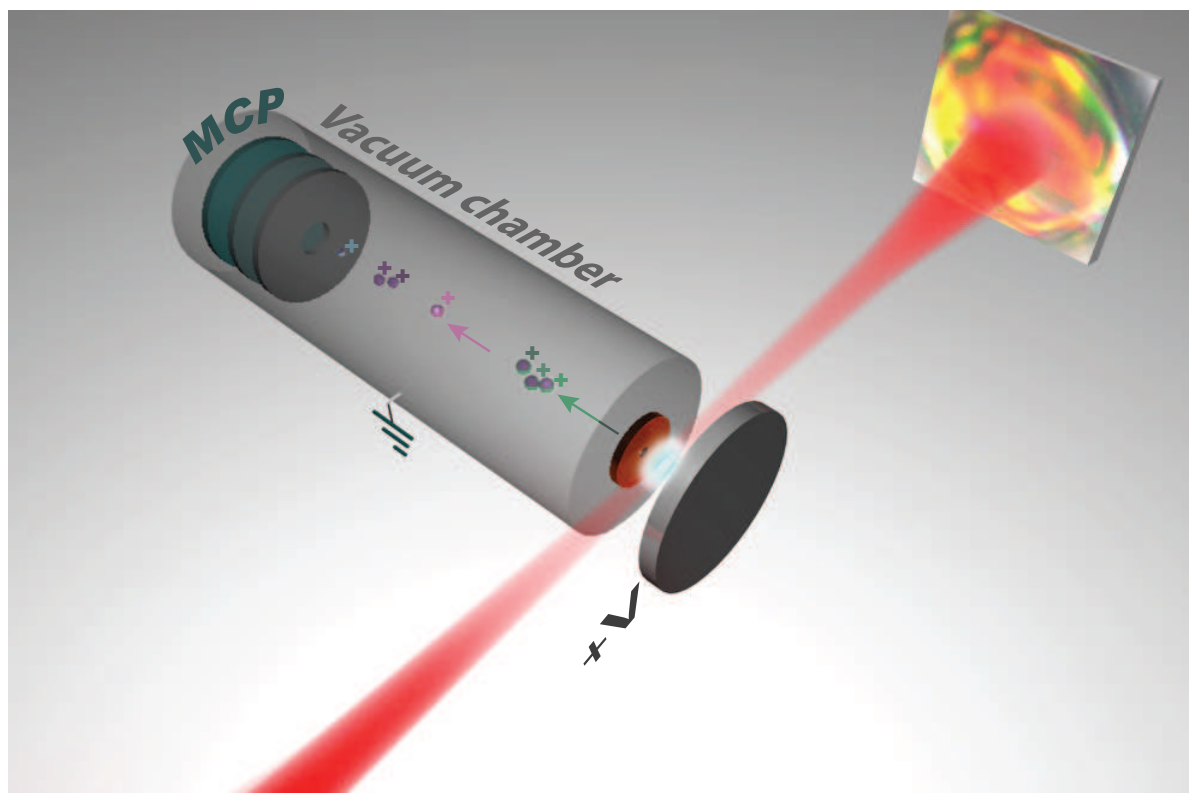}}
\caption{\label{fig1} A pulsed intense laser is focused in air or in a gas cell generating a filament. The produced ions are repelled by a positive voltage plate and driven towards the TOF spectrometer. A pinhole with a diameter of 30\,$\mu$m aligned with the plasma allows the ions to enter the high-vacuum chamber of the TOF spectrometer and reach a microchannel plate (MCP) detector after a free flight of 30\,cm. The distance from the filament to the pinhole is 2\,mm, and from the filament to the repeller plate 10\,mm. A supercontinuum is also generated in the filament.}
\end{figure}

For this work we have used a Ti-Sapphire laser system which delivers pulses of 30\,fs, up to 6\,J of energy per pulse, 800\,nm of central wavelength, a beam diameter of $10$ cm, and a repetition rate of 10\,Hz. We have attenuated the energy per pulse to 30\,mJ in order to work at atmospheric pressure conditions after compression. This laser is part of VEGA laser, which consists in three different exits with peak powers of 20\,TW, 200\,TW, and 1\,PW each.  In Fig.\,\ref{fig1} we show a sketch of the experimental setup. The laser pulses generated by the system previously described with a beam diameter of around 10\,cm (top hat profile) were focused by the combination of a spherical mirror and a plano-convex lens with focal lengths of 1\,m and 20\,cm respectively. The effective focal length of the whole system was 28\,cm. In these conditions, in the vicinity of the focal plane a filament is created.  In order to have access to the plasma dynamics and the different species generated once the laser-matter interaction ceases, we have constructed a modified version of a TOF spectrometer (see Fig.\,\ref{fig1}). In this apparatus, the interaction region at atmospheric pressure and the body of the spectrometer at a vacuum level lower than  $10^{-4}$\,mbar (this vacuum is required for a properly functioning of the MCP) are separated by a pinhole with a diameter of 30\,$\mu$m and a thickness of around 10\,$\mu$m. The ions generated in the filament are directed towards the pinhole by a repeller plate at a positive voltage. Once the ions enter the spectrometer they travel freely, i.e., there are no further acceleration stages, towards the MCP. To avoid a charge accumulation in the body of the TOF spectrometer, the outer structure must be properly connected to ground.  The signal from the MCP is collected and integrated in an oscilloscope triggered with the laser by a photodetector signal.  The integration time for the experimental data shown in this work was approximately five minutes due to the small size of the entrance pinhole and the low repetition rate of the laser.  Although a bigger pinhole would make measurements faster, it would decrease the vacuum level to a non-safe value for the correct function of the MCP, producing unwanted sparks in the detector that would ruin any useful data.  The alignment of the laser filament with respect to the pinhole was carried out maximising the UV light detected by the MCP at time zero.

\section{Experimental results and discussion}
\begin{figure}[ht!]
\resizebox{1\textwidth}{!}{\includegraphics{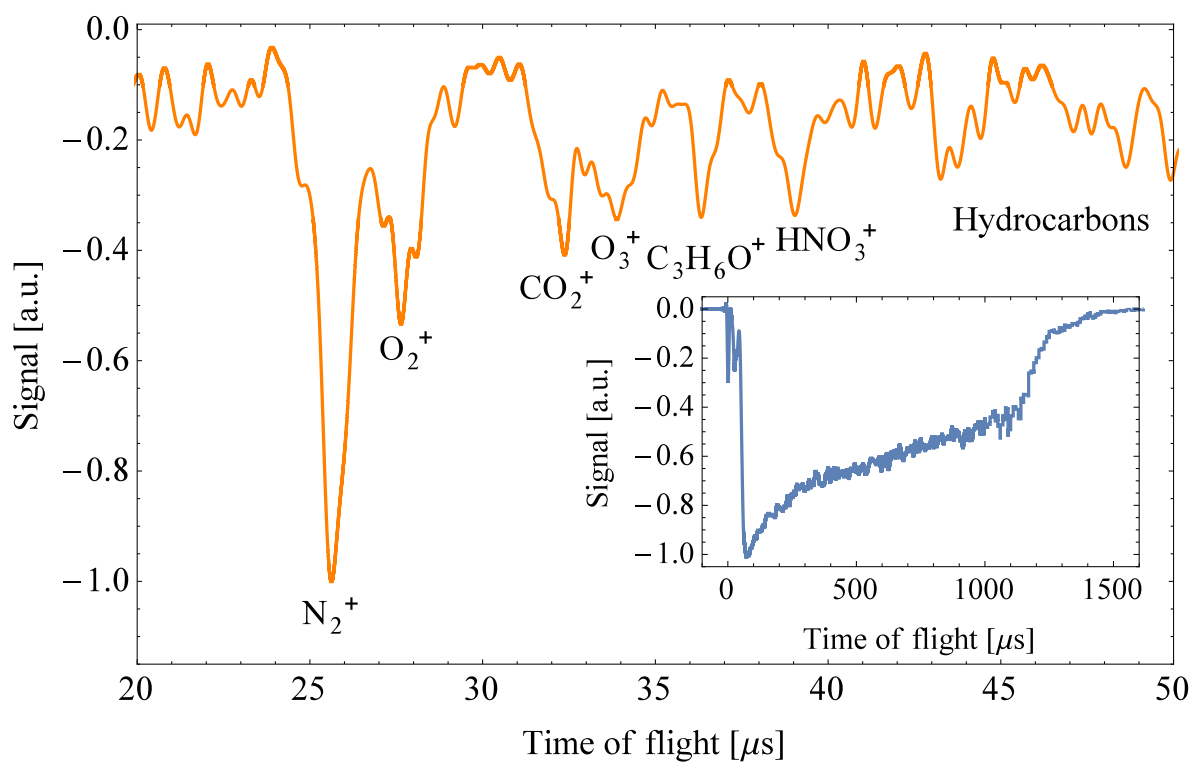}}
\caption{\label{fig2} Time of flight spectrum of a laser filament in air at atmospheric pressure. The voltage of the repeller was set to V=1\,kV.}
\end{figure}

Figure\,\ref{fig2} shows the ion spectrum when the filament was produced in air at atmospheric pressure. The data were obtained by accumulating the signal by several minutes, and applying to the row data a FFT filter for removing the fast frequencies produced by electronic noise. Three different features can be easily identified (see inset): a clear signal at time zero, a resolved ion structure for about 50\,$\mu$s, and an unresolved continuous signal of about 1\,ms of duration. The signal at time zero was produced by the UV light emitted by the induced plasma. It is important to notice that at longer times no light produced by the plasma relaxation was measured. As indicated before, we used this signal to correctly align the filament with respect to the pinhole. An explanation of the ion signals requires a more detailed analysis.  Let us focus first on the resolved spectrum (see Fig.\,\ref{fig2} main graph). Taking into account that air is mostly composed by N$_2$, it is plausible to identify the largest peak of the spectrum to this molecule and accordingly identify the rest of species.  Thus, we can detect in Fig.\,\ref{fig2} the different species of air; mainly N$^+_2$, O$^+_2$, and CO$^+_2$. It is important to notice that a quantitative analysis is not possible because at this laser intensities molecular photofragmentation is quite likely. This fact is confirmed by the appearance of new compounds like O$^+_3$ and HNO$^+_3$ which are usually not present in air. The ions found at a time of flight of around 36\,$\mu$s or 50\,$\mu$s, corresponding to a mass of 57\,amu and 101\,amu respectively, are typically assigned in mass spectrometry to a hydrocarbon chain and its photofragments, and are possibly produced by an unwanted contamination. The large peak at 36\,$\mu$s can also be tentatively assigned to some unwanted traces of acetone used to clean the optics and/or the apparatus. The structure from around 50\,$\mu$s to 1.2\,ms in Fig.\,\ref{fig2} inset shows an unresolved spectrum of ions either accelerated towards the spectrometer at the same time but with an initial kinetic energy distribution extremely large, or accelerated at very different times ones with respect to the others. To address correctly this question we first need to pay attention to the origin of both ion signals. It can be also argued that the observed peaks have a contribution from the ablation of the metallic plate holding the pinhole. We ruled out this argument because we could not identify any metallic ions like for example Al$^+$ or Cu$^+$, and different pinholes (different materials) provided the same signal.

\begin{figure}[ht!]
\resizebox{1\textwidth}{!}{\includegraphics{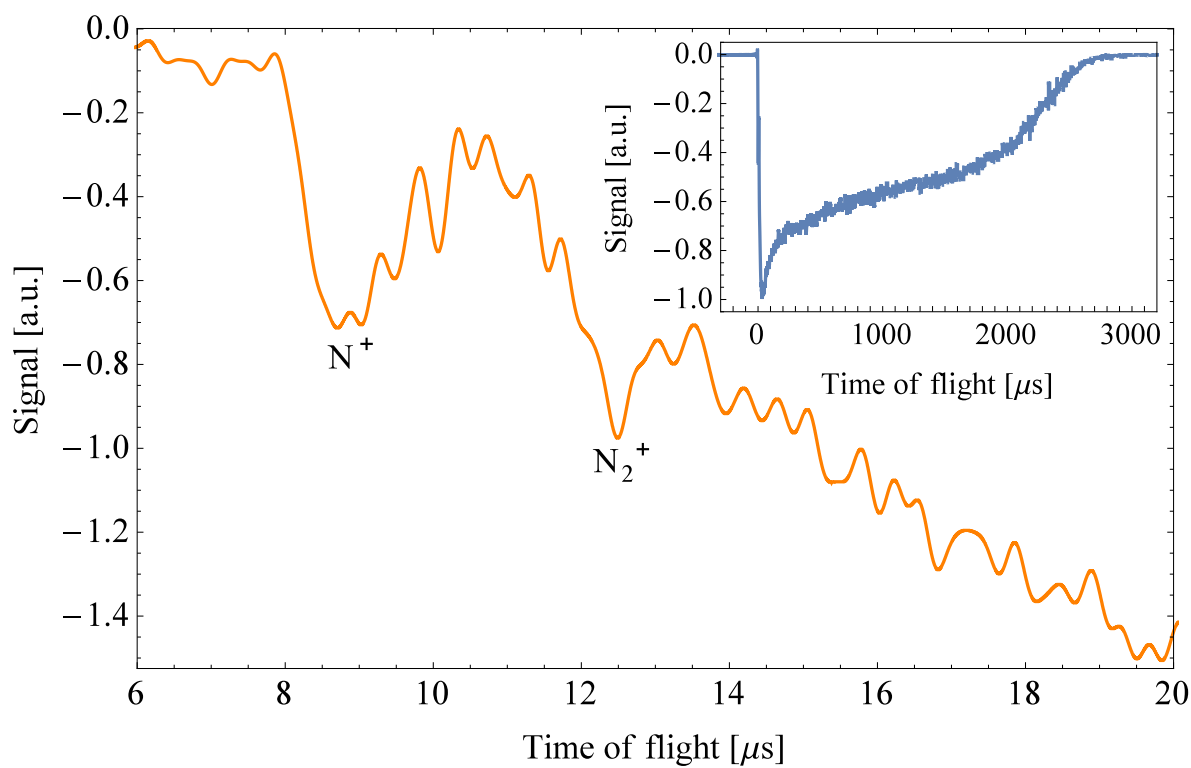}}
\caption{\label{fig3} Time of flight spectrum of a laser filament in N$_2$ at atmospheric pressure. The voltage of the repeller was set to V=2.5\,kV.}
\end{figure}

For a better understanding of the origin of the ion signals, Fig.\,\ref{fig3} shows the mass spectrum of a filament generated in pure nitrogen at a pressure of 1\,atm. For this purpose we built a methacrylate cell that could be hold in front of the pinhole with orifices for the entrance and output of the laser and for the filling gas. Similarly to Fig.\,\ref{fig2} we can first see the ion peaks corresponding to N$^+_2$ and N$^+$ (produced by the photofragmentation of the parent molecule) followed by an unresolved mass spectrum. According to the experimental results, it is plausible to think that ions in the resolved part of the spectra have their origin in the energy reservoir that surrounds the core of the filament which sustains its propagation (see for example \cite{Mysyrowicz07} and reference therein). The laser intensity reached in this reservoir is much lower than in the filament core but it is still sufficient to produce photoionization and photofragmentation of molecules. Another possible ionization mechanism is electron-impact ionization. In the first stages of the filament formation, electrons are heated up to a free-electron temperature of the order of 10$^4$-10$^5$\,K being the thermal equilibrium reached in the order of tens of nanoseconds once the laser interaction has ceased. This relaxation is mediated by elastic and inelastic collisions that produce new ions in the surrounding molecules. Attending to Fig.\,\ref{fig2} it is interesting to notice that during the relaxation path of the primary air molecules, new compounds like nitric acid (HNO$_3$) or ozone (O$_3$), energetically more stable than air compounds, are formed. The presence of these molecules indicates not only that in the energy reservoir a major ionization and photofragmentation is taking place but also a highly collisional environment. The final relaxation products of laser filamentation in air is attracting a huge interest lately by their role in the control of atmospheric processes using laser filaments. For example, the role of HNO$_3$ in induced snow formation by laser filamentation was recently acknowledge by the group of Prof. S.L. Chin (see for example \cite{Chin14}).

\begin{figure}[ht!]
\resizebox{1\textwidth}{!}{\includegraphics{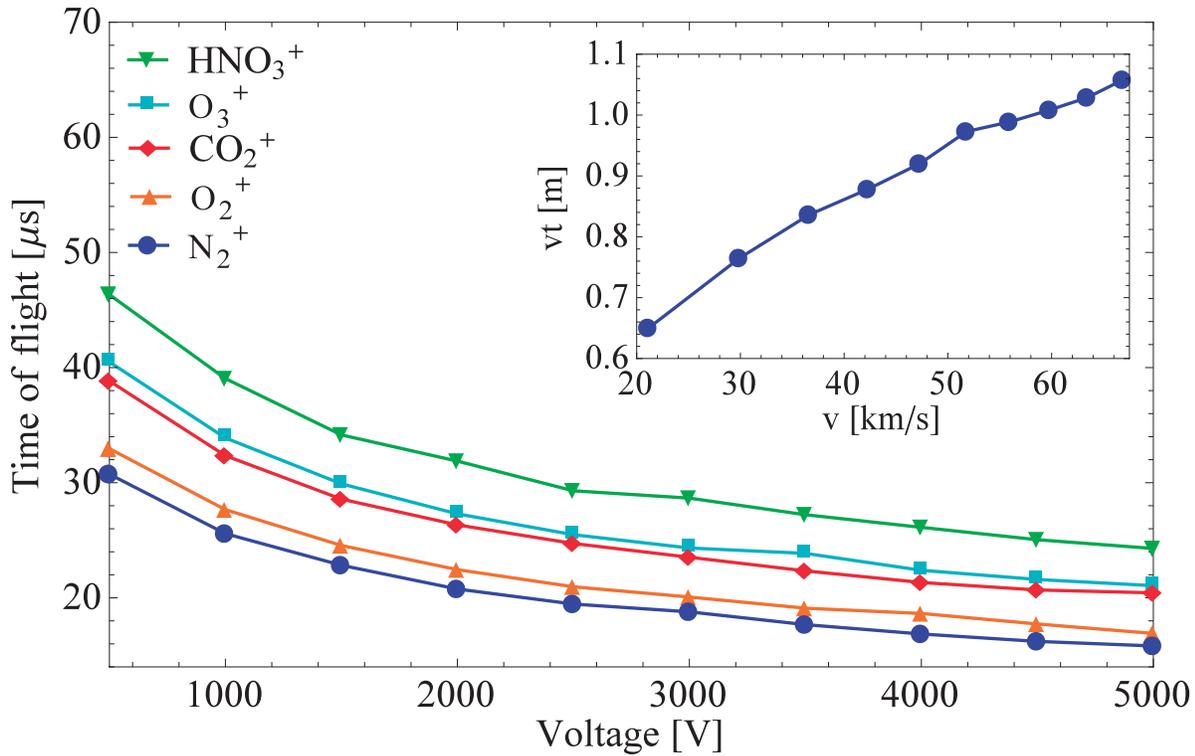}}
\caption{\label{fig4} Main graph: time of flight versus voltage applied in the repeller for different species in experimental conditions similar to Fig.\,\ref{fig2}. Inset: expected distance travelled by N$_2^+$ molecules versus its expected velocity.}
\end{figure}
 
Another fact that must be discussed is the rather long time of flights measured for the different species when compared with an ordinary TOF at high vacuum conditions that range in the order of a few microsecond. The main graph in Fig.\,\ref{fig4} shows the time of flight versus voltage applied in the repeller plate for different ion species. All the plots follow the same tendency t$\propto$V$^{-\frac{1}{2}}$ indicating that although the ions are generated in the vicinity of a plasma at atmospheric conditions, they suffer a first acceleration stage. Figure\,\ref{fig4} inset shows the expected distance of N$^+_2$ ions versus the expected velocity for different voltages of the repeller plate. This velocity is calculated taking into account the acceleration provided by the repeller voltage plate. In an ideal time of flight spectrometer this curve must be horizontal because all the ions travel the same distance. In that case the intersection with the abscissa axis is the length of the spectrometer. Here we observe that the flown distance increases linearly with the velocity. This indicates that there is some sort of mechanism that delays the particles in their trip to the detector. The magnitude of this delay is given by the slope of the curve being in this case of around 8.5\,$\mu$s. We attribute this delay time to the multiple collisions that the generated ions suffer in the atmosphere just before the pinhole entrance to the spectrometer. According to this, collisions explain partly the difference between the measured and theoretical time of flights. To get a complete picture of the problem we also need to take into account that the filament core once the laser is gone, i.e., the plasma left behind, behaves like a dielectric material. Therefore, the plasma gets polarized by the voltage difference between the repeller plate and the grounded spectrometer, having as a consequence the cancelation of the electric field. Accordingly, the effective acceleration voltage is extremely low producing the large observed time of flights. It is true that the plasma expands, and consequently its density is reduced, but the time scale of this process is much slower than the time of flight of the ions generated in the reservoir region. This idea links directly with the origin of the unresolved ion signal seen in Fig.\,\ref{fig2} and Fig.\,\ref{fig3} insets. We attribute this unresolved part of the measurements to the ions generated in the filament core. Once the interaction ceases a plasma with an electron density in the order of 10$^{16}$\,cm$^{-3}$ is formed. Although plasmas with this electron densities are underdense the electromagnetic forces between ions and electrons are sufficient to hold the species together preventing any significant leak towards the spectrometer. In fact, as it was discussed above, there is an effective charge separation in the plasma induced by the external acceleration voltage. Once the plasma expands the density drops and thus the forces among the plasma species are reduced. The experimental data indicates that this process is gradual producing a continuous leak of ions and therefore an unresolved ion signal.

\section{Numerical simulations}

To justify from a theoretical point of view the conclusions drawn from the experimental data, we developed a one dimensional PIC (particle in cell, see for example \cite{Langdon85})  code to simulate the behaviour of a plasma under the influence of an external electric field. More concretely, we implemented a plasma with an electronic density of 10$^{20}$\,m$^{-3}$ in a square profile of 20\,$\mu$m of width at t=0 in vacuum. The initial temperature was set to 1\,eV and the applied external field to -10$^7$\,Vm$^{-1}$. It is important to recognize that our choice of the parameters and the external field represents an extreme situation that favors the charge separation in plasmas for a better visualization of the results. In laser-based filaments the density and the temperature are of the order of 10$^{22}$\,m$^{-3}$ and 0.5\,eV respectively \cite{Chin11,Rudolph11}. Nevertheless the choice of parameters does not affect the Physics that underlies in this process. As PIC parameters we selected a time step of 50\,fs, a spatial cell of 0.01\,$\mu$m, and a macro particle size of 10$^4$ (real particles per computational particle). With these characteristics the plasma presents a Debye length of 0.74\,$\mu$m and a plasma frequency of  $\sim$9$\cdot$10$^{10}$\,s$^{-1}$. 
 
 \begin{figure}[ht!]
\resizebox{1\textwidth}{!}{\includegraphics{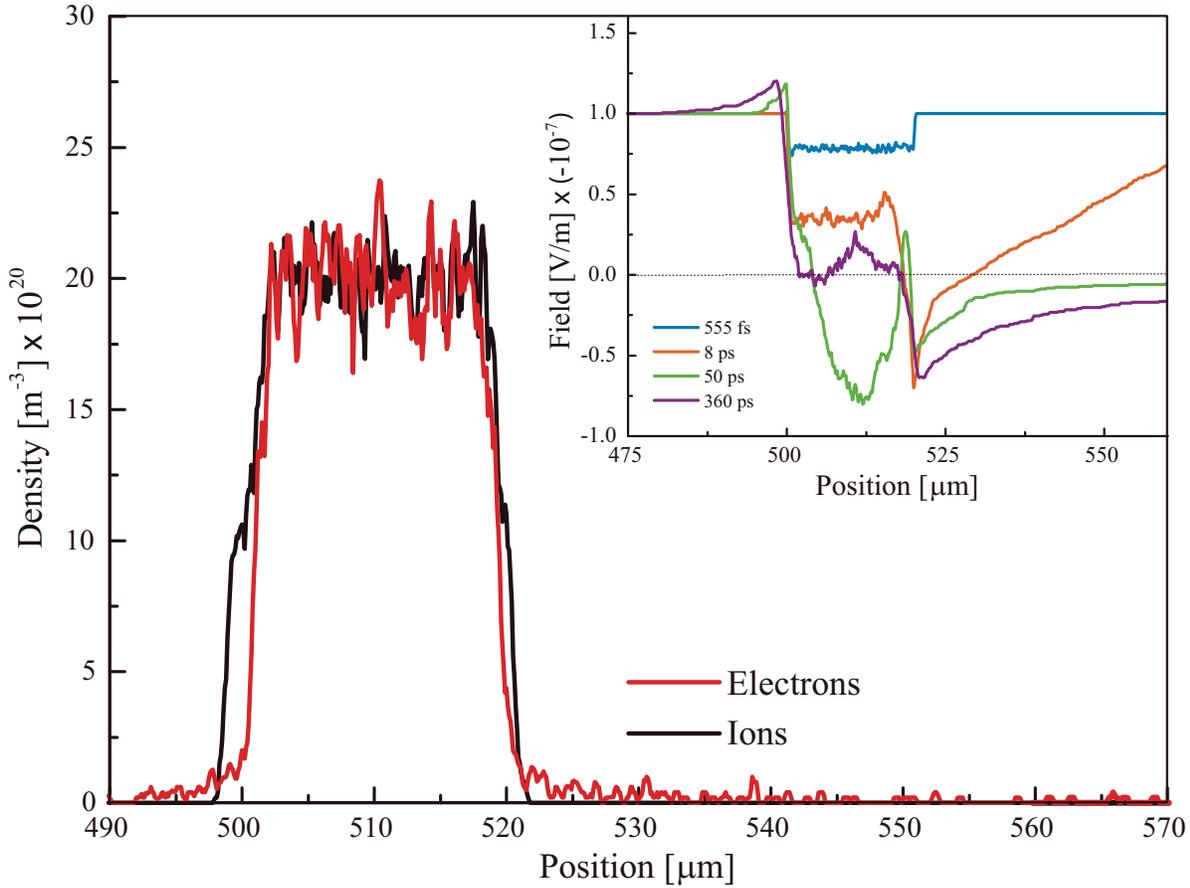}}
\caption{\label{fig5} Main graph: ion and electron density at 360\,ps of time evolution. Inset: total field, i.e., external applied field plus internal field generated by the plasma polarization, for different times of evolution. The parameters of the simulation were chosen for the sake of the argument: computational time, better visualization, etc. No relevant differences are expected with another set of parameters.}
\end{figure}
 
Figure\,\ref{fig5} main graph shows the ion and electron density at 360\,ps of time evolution. It can be clearly seen an effective charge separation in the plasma (notice the separation of electrons and ions at the edges) producing its polarisation and consequently the cancellation of the external electric field. For a better understanding of this phenomenon the inset of Fig.\,\ref{fig5} shows the total field, i.e., the sum of the external field plus the internal one created by the charge separation, as a function of the position for different evolution times. Even at the first stages of the evolution the plasma becomes quickly polarised under the influence of the external field. The total field inside the plasma oscillates around zero at the plasma frequency. It must be noticed that at the first stages of the dynamics, before the plasma gets polarised and there is an effective charge separation, some electrons can escape from the plasma deviating it slightly from the electrical neutrality. This non-neutrality of the plasma causes the field to be shielded not only inside its boundaries but also outside it, following the escape direction of the electrons (right side in Fig.\,\ref{fig5}). On the opposite side (from the plasma to the left in the graph) the field remains at its initial value of -10$^7$\,Vm$^{-1}$. This behaviour stays constant until the ions expand significantly, reducing the density and thus increasing the Debye length to scales of the order of the experiment boundaries. Finally, it is interesting to pay attention to the  sheaths at the edges of the plasma in Fig.\,\ref{fig5}. These sheaths give rise to field gradients at the plasma boundaries that extend for several tens of Debye lengths.    

\section{Conclusions}
In conclusion, in this work we have studied by mass spectrometry techniques the ions generated by a laser filament. Our original mass spectrometer design allowed us to distinguish the ions generated within the filament energy reservoir, characterized by a clear mass spectrum, and the ones coming from the filament core, characterized by a continuous one. The sensitivity of the device also permitted us to study the photofragmentation taking place in the plasma and surroundings. Furthermore, analyzing the form and characteristics of the mass spectra for both components, we obtained valuable information about the plasma dynamics and relaxation channels. For example, final relaxation products with huge importance for filament based atmospheric applications like O$_3$ and HNO$_3$ have been clearly observed. In our opinion this work and the new developed apparatus pave the way for further studies on plasma dynamics and relaxation using mass spectrometry techniques. An example of a further development is the use of high voltage switch to pulse the repeller voltage at a controllable temporal delay with respect to the laser pulse. Operating in this way, it is possible to obtain snapshots at different moments of the relaxation dynamics obtaining a very valuable information of the involved relaxation processes. 

\section{Acknowledgements}

This work has been possible by the support from Ministerio de Educaci\'on Cultura y Deporte (F. Valle Brozas studentship FPU AP2012-3451), Ministerio de Econom\'ia y Competitividad of Spain (SIGMA project IPT-2011-1137-310000 and FURIAM project FIS2013-47741-R), Junta de Castilla y Le\'on (CLP281U14 project), and the EC's Seventh Framework Programme LASERLAB-EUROPE III (grant agreement 284464). M.S.A. also thanks the Junta de Castilla y Le\'on and Fondo Social Europeo for the economical funding.

\end{document}